\documentclass{aa}
\usepackage{epsfig,amssymb}
\usepackage{graphicx}
\begin{document}
%\thesaurus{06(06.03.2; 06.13.1; 03.13.6; 02.03.1)}
%
\title{Quiet Sun coronal heating: analyzing large scale
magnetic structures driven by different small-scale uniform
sources.}

\subtitle{}

\author{O.~Podladchikova \and T.~Dudok~de~Wit
\and V.~Krasnoselskikh \and B.~Lefebvre}

\offprints {O.~Podladchikova,\\ email:epodlad@cnrs-orleans.fr}

\institute{ LPCE/CNRS, 3A av. de la recherche scientifique,
            45071 Orl\'eans, France}
\date{Received 15 August 2001; accepted }

\authorrunning{O.~Podladchikova et al.}
\titlerunning{Analyzing large scale
magnetic structures...}

\abstract{ Recent measurements of quiet Sun heating events by
Krucker \& Benz (\cite{Krucker98}) give strong support to Parker's
(\cite{par}) hypothesis that small scale dissipative events make
the main contribution to the quiet heating. Moreover, combining
their observations with the analysis by Priest et al.
(\cite{pr1}), it can be concluded that the sources driving these
dissipative events are also small scale sources, typically of the
order of (or smaller than) 2000 km and the resolution of modern
instruments. Thus arises the question of how these small scale
events participate into the larger scale observable phenomena, and
how the information about small scales can be extracted from
observations. This problem is treated in the framework of a simple
phenomenological model introduced in Krasnoselskikh et al.
(\cite{KPL}), which allows to switch between various small scale
sources and dissipative processes. The large scale structure of
the magnetic field is studied by means of Singular Value
Decomposition (SVD) and a derived entropy, techniques which are
readily applicable to experimental data.
}

\maketitle

%%%%%%%%%%%%%%%%%%%%%%%%%%%%%%%%%%%%%%%%%%%%%%%%%%%%%%%%%%%%%%%
\section{Why small-scale sources ?}

The anomalously high temperature of the solar corona is still a
puzzling problem of solar physics, despite the considerable
theoretical and experimental efforts involved for a long time
(e.g. Priest et al., \cite{pr1}; Einaudi \& Velli, \cite{Einn}).
 Since the energy release in the largest heating events (flares
and microflares) does not supply enough power to heat the Corona,
the statistical behavior of smaller-scale and less energetic but
much more frequent events is an essential feature of the problem,
as was conjectured some time ago by Parker (\cite{par}).

An important result that supports Parker's hypothesis was reported
by Krucker \& Benz (\cite{Krucker98}) who have found from the
Yohkoh / SXT observations, assuming that the flaring region has a
constant height, that the energy probability density has the form
of a power law in the energy range $10^{24}$--$10^{26}$ ergs with
an exponent about $-2.59$. Such an exponent less than $-2$
indicates that heating takes place in small scales, while on the
contrary an exponent greater than $-2$ indicate that large scale
phenomena play a dominant role. The conclusion of Krucker \& Benz
was confirmed by Parnell \& Jupp (\cite{Parnell00}), who estimated
the exponent to be between $-2$ and $-2.1$ making use of the data
of TRACE and supposing that the height varies proportionally to
the square root of the area. The relatively steep slope of the
energy distribution found by these authors also suggests that the
smallest flares contribute essentially to the heating. Mitra \&
Benz (\cite{Mitra01}) have discussed the same observations as
Krucker \& Benz but supposing the height variations similar to
Parnell \& Jupp, they have shown that the exponent becomes a
little larger then previous estimate but still is smaller than
minus two.

Making use of the multi-wavelength analysis Benz \& Krucker
(\cite{Benz99}) have shown that energy release mechanisms are
similar in large scale loops and in the faintest observable
events. They have also noticed that the heating events occur not
only on the boundaries of the magnetic network but in the
interiors of the cells also. Priest et al. (\cite{Priest98}),
comparing model predictions for the plasma heating in the magnetic
loop due to the distributed energy source with observations led to
the conclusion that the heating is quasi-homogeneous along the
magnetic loop. This means that the heating process does not occur
in the close vicinity of the foot points but rather in the whole
arc volume. Putting together these facts, it follows that the
characteristic spatial scale of the magnetic field loops which
supply the magnetic field dissipated is of the same order as the
characteristic scale of the dissipation. Thus we may conclude that
not only the dissipative process, but also the energy sources have
small characteristic length. It also results from these
observations that the sources are distributed quite homogeneously
in space.

Hence it is important to discuss the role and properties of
sources and dissipative processes in the framework of simplified
models. Such an approach allows to study the correspondence
between large scale magnetic field properties and characteristics
of the small scale (eventually smaller than the experimental
resolution) characteristics of sources and dissipative events,
which may become a basis of the analysis of experimental data,
helping to explain the nature of the physical phenomena underlying
the observations.

A phenomenological model allowing for different sources and
physical dissipation mechanisms was proposed in Krasnoselskikh et
al. (\cite{KPL}), and their effect on the temporal statistics of
the total dissipated energy were studied. This model is briefly
discussed in the next section. To study spatial properties of the
magnetic fields and dissipative events, detailed statistical tests
applicable both to simulations and experimental data are required.
Tools such as the magnetic field entropy and extraction of the
most energetic and large scale spatial/temporal eigenmodes by
means of Singular Value Decomposition (SVD) are described in
section III. Their application to our model and their ability to
discriminate between various sources and dissipative mechanisms
are discussed in section IV, and the final section proposes a
review and a critical discussion of the results.

\section{Small-scale driving and dissipation}

Various phenomenological models of flare-like events and
cooperative phenomena in the corona have been considered in the
literature (e.g. Lu \& Hamilton, \cite{Lu-H}; Vlahos et al.,
\cite{vlah2}; Georgoulis et al., \cite{Georg4}), mostly relying on
the notion of Self-Organized Criticality (Jensen, \cite{Jensen}).
 Such models usually exhibit infinite-range spatial correlations,
 and due to the tenuosity and localization of the driving do not
 provide an appropriate framework for our purpose.

 Instead, we shall use the model introduced in
(Krasnoselskikh et al., \cite{KPL}; Podladchikova et al.,
\cite{PKL}) which allows for a driving more distributed in space
and dissipative processes relevant to heating studies. As usual,
the model represents a simplification of the magnetohydrodynamic
induction equation
\begin{equation}
\frac{\partial\mathbf{B}}{\partial t}
=\mathbf{\nabla}\times(\mathbf{u}\times\mathbf{B}) +
\mathrm{dissipative\ term.} \label{B}
\end{equation}
The turbulent photospheric convection randomizes in some sense the
first term of the right hand side, which can be replaced by
various source terms with specified statistical properties. The
dissipative terms may take into account different effects such as
normal and anomalous resistivity or magnetic reconnection, which
in general depend on the current density and magnetic field
configuration (their meaning and differences between the two in
this context are discussed at length in Krasnoselskikh et al.
(\cite{KPL}).

The model is two-dimensional, the magnetic field being
perpendicular to the grid, with periodic boundary conditions. A
discrete description of the magnetic field in term of cells is
proposed, while the currents are computed from
\begin{eqnarray*}
\left(
\begin{array}{c}
   j_x \\ j_y
\end{array} \right) = \frac{1}{\delta} \left(
\begin{array}{c}
B\left(x,y\right)-B\left( x,y+\delta \right) \\
B\left(x+\delta,y\right)-B\left( x,y\right),
\end{array}
\right)
\end{eqnarray*}
where $\delta$ is cell length ($\delta=1$ in the following). The
currents can be considered as propagating on the border between
the cells, and satisfy Kirchoff's law at each node.

\begin{figure}
\centering
\includegraphics[width=8.8cm]{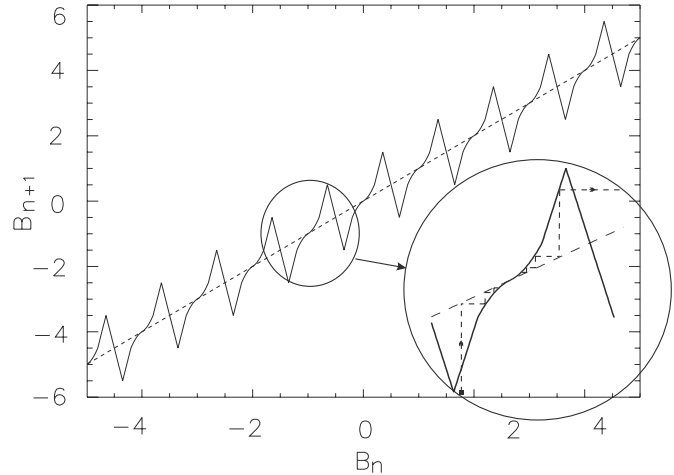}
\caption{Graphical representation of the Geisel map (solid line).
The fixed points of the map correspond to the intersections of the
graph with the straight line $B_{n+1}=B_n$ (dashed line).}
\label{Geiselmap}
\end{figure}

As discussed in the introduction, one may suppose that the source
term representing the magnetic energy injection has a
characteristic spatial comparable with that of the dissipative
processes. Hence source terms, mimicking the magnetic energy
injection from the turbulent photosphere, are assumed to have a
vanishing average, and act in each cell of the grid at each time
step. Three different types with different statistical properties
 are considered:
\begin{itemize}
\item{\bf Random sources.}
The simplest source is given by independent random variables
$\delta B$ in the set $\{-1,0,1\}$, acting in each cell. Such a
source can be made dipolar by dividing the grid into two parts
where random numbers are chosen from the similar set that have
however, positive and negative mean values, respectively, for each
of these parts.
\item{\bf A chaotic source.}
Turbulence is certainly not a completely random process, and some
of its aspects are enlightened by deterministic models. The source
in each cell evolves independently of others according to
\[
\delta B_{n+1}=1-2(\delta B_n)^2,
\]
where $\delta B\in[-1;1]$,
which is an instance of the logistic (Ulam) map, well known for its chaotic
properties.
\item{\bf Geisel map source.}
The source term may depend on the value of $B$ itself. When the
dissipation is absent, the magnetic field in each cell evolves
according to a map
\[
B_{n+1}=f(B_n).
\]
We used the map introduced by Geisel and Thomae, (\cite{Geisel}),
hereafter called Geisel map (see Fig.~\ref{Geiselmap}), whose
marginally stable fixed points are responsible for the anomalous
diffusion exhibited by this map
\[
\langle B^2\rangle\propto t^\alpha,\quad\alpha<1.
\]
It is generally expected that magnetic field lines in a turbulent
plasma exhibit a subdiffusive behavior, which is, however, more
complex than described above.
\end{itemize}

% put this text after this  figure

It is worth pointing out that all the source evolve in time in
each cell independently of other cells, and that interaction
between the fields in neighboring cells are only related with the
dissipation effects.

The dissipation provides the conversion of magnetic energy into
particle acceleration and thermal energy, and in our model
provides the coupling between the magnetic field elements.
Dissipative processes are most important where a current sheet
carrying strong current density has formed. Neglecting
resistivity, which is small in the corona, one is left with
various instabilities of magnetic field configurations which can
provide the dissipation. We consider two of them:
\begin{itemize}
\item {\bf Anomalous resistivity},
which arises from the development of certain instabilities such as
modified Buneman when the electric current exceeds a certain
threshold in collisionless plasma. In our model the currents are
simply annihilated whenever they exceed a certain threshold,
\[
|j|\geq j_{\max}.
\]
\item {\bf Reconnection},
for which we impose which in our model the additional requirement
that the magnetic field has a configuration favorable to a
X-point. Hence, the following two conditions should be satisfied
simultaneously:
\begin{eqnarray}
|j|=|B-B'| & \geq & j_{\max}, \nonumber \\ B\cdot B' & < & 0,
\label{BB'}
\end{eqnarray}
The new condition results in the existence of currents that can
strongly overcome the critical value.
\end{itemize}
The difference underlined here is mainly that reconnection
represents a change in equilibrium, from one topology (here a
X-point) to another, while anomalous resistivity does not require
any particular topology and thus may also act in cell interiors
and not only at boundaries. Another important difference is that
anomalous resistivity provides Joule-like heating, while
reconnection yields accelerated outgoing flows and thus may be
associated to non-thermal radiation.

 When the current
is annihilated, magnetic field values in both neighboring cells,
$B$ and $B'$, are replaced by $1/2(B+B')$, so the density of
magnetic energy dissipated in a single event is given by
\[
\Delta E = \frac{1}{4}\,(B-B')^2 = \frac{1}{4}\,j^2 \gtrsim
\frac{1}{4}\,j_{\max}^2.
\]
The procedure modeling the dissipation of currents is the same for
both anomalous resistivity and reconnection. On each time step,
the currents satisfying the dissipation criterion are dissipated
till all the currents become subcritical (or have the same sign
for the case of reconnection). Then, we proceed to the next time
step and switch on the source. Indeed, dissipative processes are
supposed to be faster than the driving ones. The total dissipated
energy is calculated as a sum over all the dissipated currents for
the time step considered.

In Podladchikova et al. (\cite{PKL}) and Krasnoselskikh et al.
(\cite{KPL}), the influence of the dissipative processes and
source terms on the statistical properties of the dissipated
energy was studied. The dissipation was shown to have a
significant influence on the statistics of dissipated energy.
Indeed the reconnection mechanism was shown to yield the strongest
deviation from Gaussian distribution in the large energies.
However, the probability density of the dissipated energy was
shown to be rather insensitive to the nature of the magnetic field
sources. In this paper we would like to explore further the
dependence of the large scale magnetic field statistical
properties upon the physical characteristics of the source and
dissipation processes in the framework of our model. The aim of
this work is to investigate the possibility of getting information
about small scale magnetic energy sources making use the large
scale magnetic field. In order to do so, we study hereafter the
spatial complexity of the large scale field by means of spatial
correlations, entropy and eigenmodes (Singular Value
Decomposition).

%%%%%%%%%%%%%%%%%%%%%%%%%%%%%%%%%%%%%%%%%%%%%%%%
\section{Characterization of spatial complexity}

Spatial complexity can be characterized in many different ways
(e.g. Grassberger, \cite{Grassberger86}). Linear properties are
traditionally studied by considering the time averaged spatial
correlation function
\begin{equation}
    \label{eq:corr}
     C({\mathbf r}) = \langle B({\mathbf x},t) \;
B({\mathbf x+r},t) \rangle_{{\mathbf x},t} / \langle B({\mathbf
x},t) ^2 \rangle_{{\mathbf x},t} \ ,
\end{equation}
 where the
average is carried out over different positions and times (or
events). We have computed the characteristic decay length of this
correlation function for various sources, dissipation mechanisms,
and thresholds.

\medskip

A different approach, which is used in the framework of image
processing, is based on the Singular Value Decomposition (SVD) or
Karhunen-Lo\`eve Transform, see Golub \& van Loan
(\cite{Golub96}). For each time step, the bivariate magnetic field
intensity $B(x,y)$ can be viewed as 2D image.  We decompose this
image into a set of separable spatial modes
\begin{equation}
    \label{eq:svd}
    B(x,y) = \sum_{k=1}^{N} \mu_k \; f_k(x) \; g_k^*(y) \ .
\end{equation}
By making these modes orthogonal $\langle f_k f_l^* \rangle =
\langle g_k g_l^* \rangle = \delta_{k,l}$, the decomposition
becomes unique. The weights $\mu_k$ of these modes, also called
singular values, are conventionally sorted in decreasing order,
and are invariant with respect to all orthogonal transformations
of the matrix $B(x,y)$. In our case, the number $N$ of modes is
equal to the spatial grid size.

A key property of the SVD is that it captures large-scale
structures in heavily weighted modes, whereas patterns that are
little correlated in space are deferred to modes with small
weights.  The distribution of the singular values is therefore
indicative of the spatial disorder: a flat distribution means that
there is no characteristic spatial scale and hence, the magnetic
field should not show large-scale patterns.  Conversely, a peaked
distribution suggests that there are coherent structures (Dudok de
Wit, \cite{DdW95}).  It must be stressed that this approach is,
like the previous one, based on second order moments only, since
the spatial modes and their singular values issue from the
eigenstructure of the spatial correlation matrix of the magnetic
field.

From the SVD modes of the 2D magnetic field, one can define a
measure of spatial complexity, which is based on the SVD entropy
(Aubry, \cite{Aubry91}). Let $E_k = \mu_k^2 / \sum_i \mu_i^2$ be
the fractional amount of energy which is contained in the $k$'th
mode. The SVD entropy can then be defined as
\begin{equation}
H = - \lim_{N\rightarrow\infty}\frac{1}{\log N} \sum_{k=1}^N E_k
\log E_k \ . \label{eq:H}
\end{equation}
The maximum value $H=1$ is reached when spatial disorder is
maximum, that is when $E_k = 1/N$ for all $k$. $H=0$ means that
all the variance is contained in a single mode.

The SVD can also be used as a linear filter to extract large scale
patterns from a background with small scale fluctuations. To do
so, one should perform the SVD and then in eq.~\ref{eq:svd} sum
over the strongest modes only, to obtain a filtered magnetic
field. There is obviously some arbitrariness involved in the
identification of what we call strong modes, but the process can
be automated by using robust selection criteria, see for example
Dudok de Wit (\cite{DdW95}).

%%%%%%%%%%%%%%%%%%%%%%%%%%%%%%%%%%%%%%%%%%%%%%%%%%%%%%%%%%%%%%%%%%%%%%%%%
\section{Spatial complexity and properties of the source and dissipation}

\subsection{Spatial correlations}

\begin{figure}
\centering
\includegraphics[width=8cm]{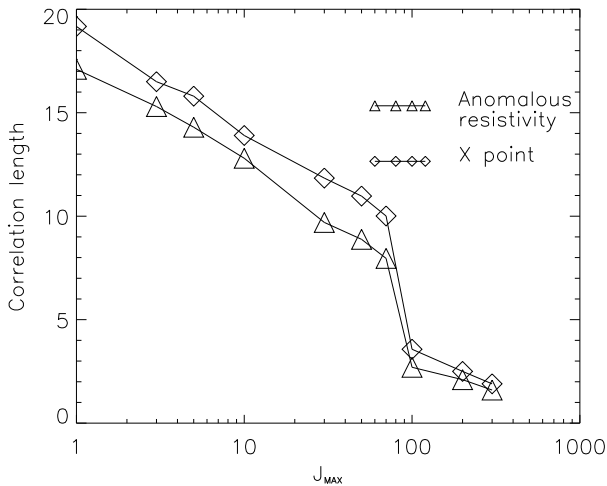}
\caption{Correlation length dependence on the threshold of
dissipation for both dissipation rules. Anomalous resistivity is
marked by triangles, reconnection by squares.} \label{Corel}
\end{figure}

\begin{figure}
\centering
\includegraphics[width=8cm]{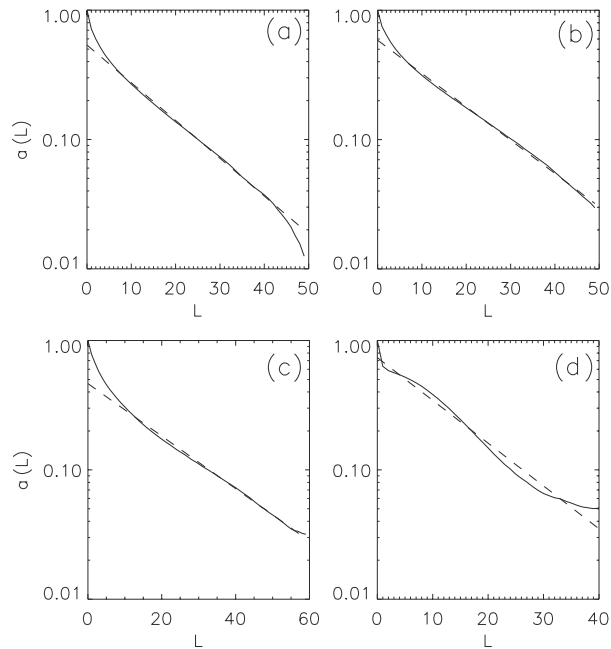}
\caption{Averaged magnetic field spatial correlation coefficient
in log-linear plot (solid lines). Dotted lines represent the best
fits by exponential function. The results are obtained for random
and subdiffusive sources with a threshold of dissipation $j_{\max}
= 1$. \textbf{a)} random source, anomalous resistivity
dissipation, correlation length $L=17$; \textbf{b)} random source,
reconnection dissipation, correlation length $L=19$; \textbf{c)}
subdiffusive source, anomalous resistivity dissipation,
correlation length $L=23$; \textbf{d)} subdiffusive source,
reconnection dissipation. } \label{Corr}
\end{figure}

\begin{figure*}
\centering
\includegraphics[width=18cm]{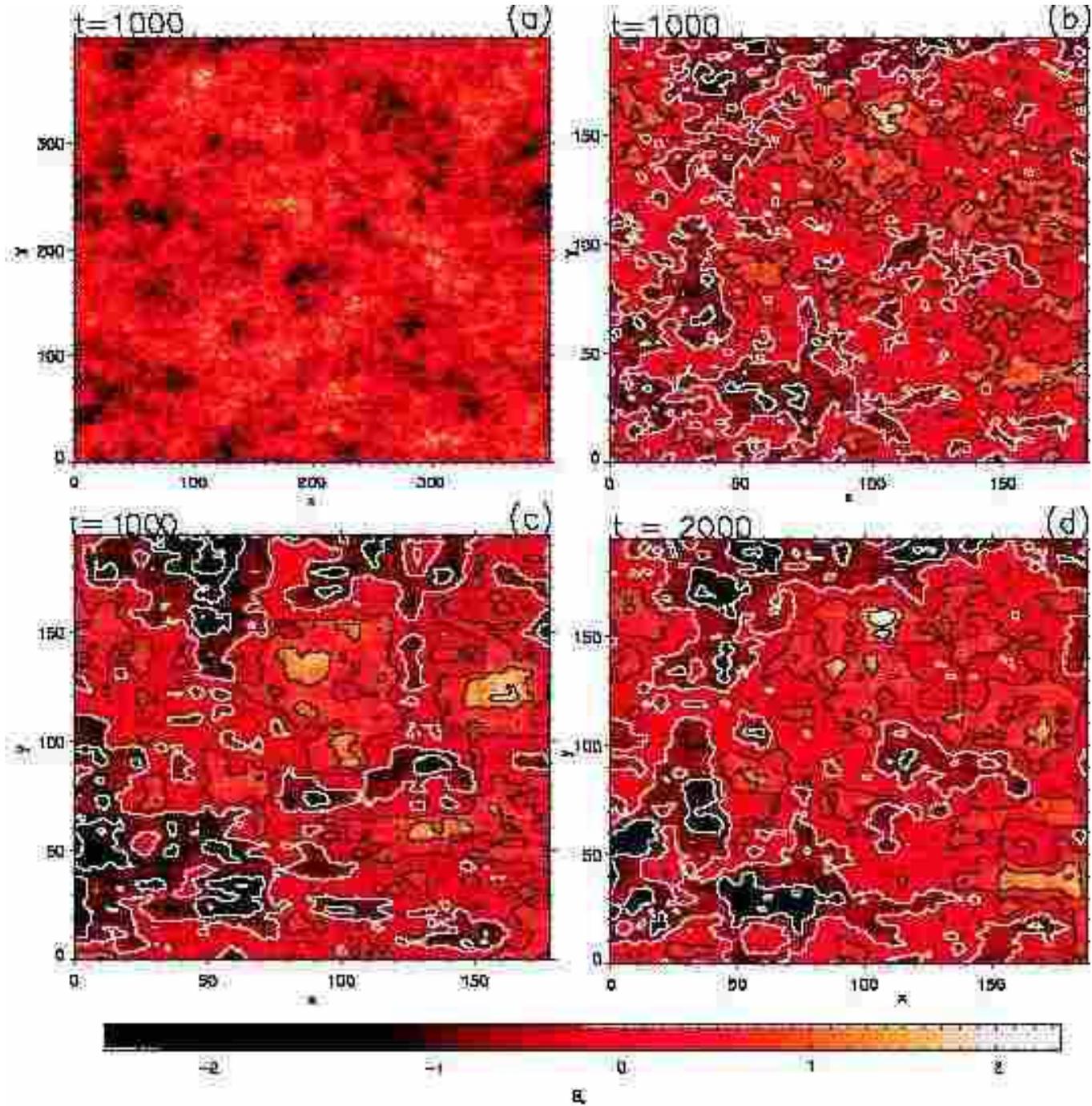}
\caption{Typical magnetic field for random source and
reconnection, with $j_{\max} = 1$. \textbf{a)} Magnetic field at
$t=1000$, with entropy $ H = 0.79 $; \textbf{b)} Zoom of the
precedent image; \textbf{c,d)} filtered magnetic field, with 20
eigenmodes, at $t=1000$ and $t=2000$
($ H = 0.807$).} \label{random}%
\end{figure*}

\begin{figure*}
\centering
\includegraphics[width=18cm]{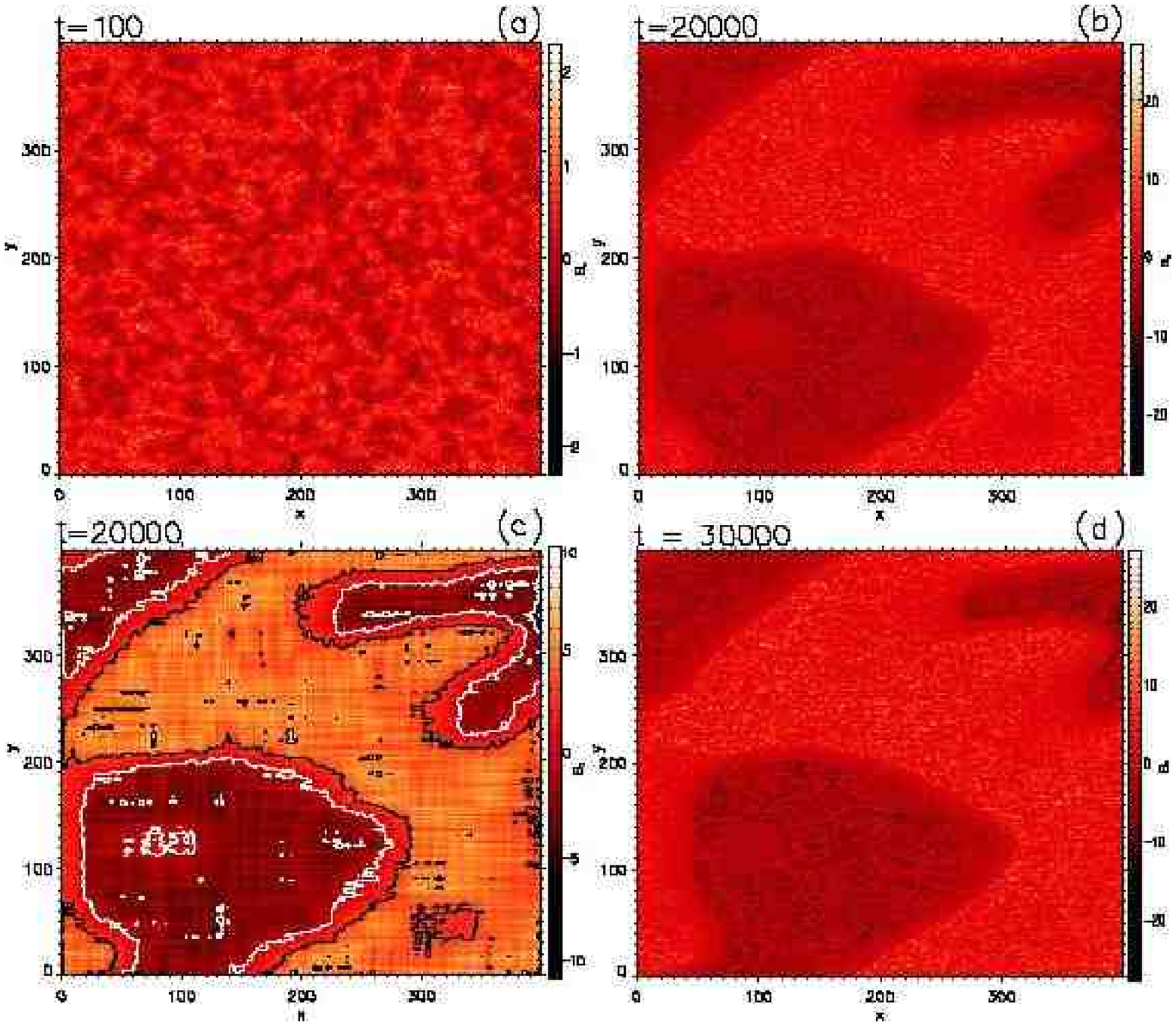}
   \caption {Magnetic field for the subdiffusive source and
   reconnection, with $j_{\max} = 1$.
 \textbf{a)} Magnetic field in a transient state, $ H = 0.73$;
    \textbf{b)}, Magnetic field in a stationary state, $ H = 0.51 $ \textbf{c)}
20 modes of the precedent image; \textbf{d)}, magnetic
field at $t = 3000$, $ H = 0.527 $}\label{geisel}%
\end{figure*}

Spatial correlation function as defined by Eq. \ref{eq:corr} were
averaged in time after the system has reached a stationary state.

For the small grids, of the order of $30\times 30$, the
correlation function decays as a power-law. It was shown in
Krasnoselskikh et al. \cite{KPL} the probability density of the
total dissipated energy also decays as a power-law in this case.
However the resemblance to a self organized critical behavior is
only an artifact of the small grid size, when the correlation
length is comparable with the grid size, and disappears for large
enough grids. Indeed, for grid sizes around $100\times 100$ (or
greater), the time averaged correlations functions have an almost
exponential decay, while the dissipated energy has a
quasi-Gaussian distribution.

In case of exponential correlation functions, a correlation length
is easily defined as the parameter $L$ such that
\[
C(r)=\exp(-r/L).
\]
We have found in our simulations that the correlation length $L$
remains almost constant when the grid size is increased above $200
\times 200$, and furthermore that $L$ is much smaller than the
linear grid size. In such a case, it is legitimate to expect that
results do not depend significantly on the grid size or boundary
conditions. In the remainder of this paper all results are
presented for a grid $400\times400$ and a small threshold is used
($j_{max} =1$, which is of the order of $\sqrt{\langle
 \delta B ^2 \rangle}$).

As shown on Fig. \ref{Corel}, for the random magnetic field source
the average correlation length is larger for the reconnection type
dissipation than for anomalous resistivity. This can be explained
by the presence of supercritical currents $>j_{\max}$ which may
exist because of condition~(\ref{BB'}). Moreover in both cases the
correlation lengths are decreasing functions of $j_{\max}$  (see
Fig.~\ref{Corel}). These results are hardly changed if the random
source is replaced by the Ulam map source (see previous section).
Only in the case of a magnetic field following the Geisel map, and
when dissipative processes are ruled by anomalous resistivity
(Fig.~\ref{Corr}c), the average correlation length is a little bit
larger than in the previous cases (see Fig.~\ref{Corr}ac). When
dissipation occurs through reconnection, the average correlation
function seems featureless (see Fig.~\ref{Corr}d) (to the least
neither exponential nor power-law), and no correlation length can
be unambiguously computed for comparison with previous results.

Thus the correlation function seems to have difficulties here in
indicating significant differences between the different
processes. However, strong differences are seen simply by visible
inspection of the magnetic field between for example the random
and Geisel map sources (Figs. \ref{random}b and Figs.
\ref{geisel}b).
 Hereafter we shall present an alternative analysis using the spatial entropy
defined in terms of SVD eigenvalues.

\subsection{Singular values and coherent spatial modes}

As discussed in the previous section, the Singular Value
Decomposition provides an orthogonal decomposition which allows to
extract coherent patterns eventually existing in the bivariate
magnetic field at a given time.

\begin{figure}
\centering
\includegraphics[width=8.8cm]{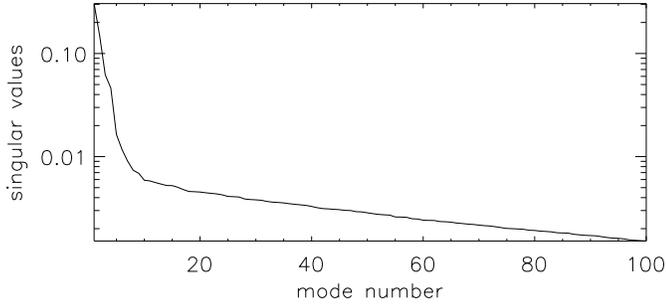}
\caption{Energy of the spatial eigenmodes of magnetic field formed
by subdiffusive source and reconnection dissipation.}
\label{podgeis}
\end{figure}

\begin{figure}
\centering
\includegraphics[width=8.8cm]{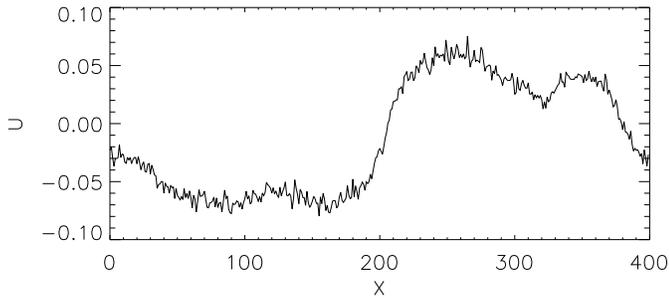}
\caption{Most energetic spatial mode $f_1(x)$ obtained by the same
Singular Value Decomposition as in Fig. \ref{podgeis}.}
\label{spmode}
\end{figure}

In our case, the distribution of singular values appears to be
significantly
 peaked (Fig. \ref{podgeis}), showing that the bi-dimensional
wavefield is dominated by a few spatial modes. For instance, the
most energetic mode ($f_1(x)$, in the notation of Eq. \ref{eq:svd}
obtained from SVD of the magnetic field for a Geisel source and
dissipation by reconnection clearly corresponds to a large scale
coherent structure (Fig. \ref{spmode}).

The fact that the most heavily weighted modes correspond to large scale
magnetic field structures is further illustrated by Figs. \ref{random} and
\ref{geisel}, by comparison of the magnetic field (Figs. \ref{random}b and
\ref{geisel}b) with an approximation of the magnetic field containing only
20 modes (Figs. \ref{random}c and \ref{geisel}c) which has a very similar
 large scale structure. In both cases, the retained modes correspond to the
fast exponential decay of the strongest singular values,
down to the turning point where the decay becomes slower (as in
Fig. \ref{podgeis}), and the following singular values are filtered out and
set to 0 before the inverse SVD is performed.
In particular, it is clear that the large scale
structure of magnetic field appearing with Geisel map sources, and not
with random sources, is retained by SVD (Fig. \ref{geisel}).

However, this analysis provides only a decomposition of the magnetic field at
a given time, and no information about the lifetime of these
structures which is quite crucial though. Monitoring the time evolution of the
system, it appears that the heavily weighted modes appear
to persist for long times, as can be seen comparing the filtered magnetic field
at two times separated by 2000 time steps (Figs \ref{geisel}c and d). Actually,
it is seen on Figs \ref{geisel}a,b,c how these structures grow from the
initially disordered state.

\begin{figure}
%\resizebox{\hsize}{!}{\includegraphics{geisel1.eps}}
\centering
\includegraphics[width=8.8cm]{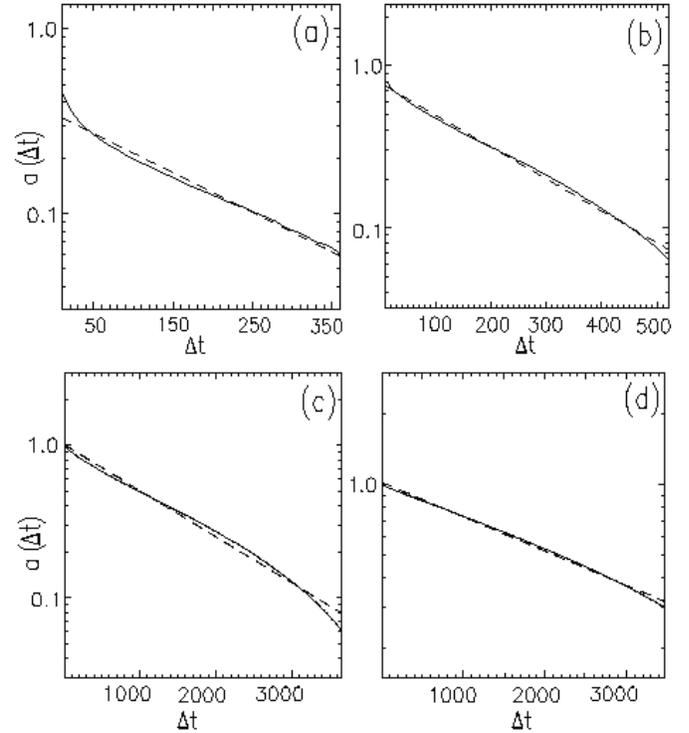}
\caption{Averaged temporal correlation function of the magnetic
field (solid lines) in log-linear plot calculated from
$4\times10^4$ times steps. Dotted lines represent the best fits by
an exponential function. The results are obtained for random and
subdiffusive sources, reconnection dissipation, with a threshold
of dissipation $j_{\max} = 1$. \textbf{a)} random source,
correlation time $\tau=202$; \textbf{b)} the same as previous, but
only for the first 20 modes, $\tau=220$; \textbf{c)} subdiffusive
source, $\tau=1435$; \textbf{d)} the same as (c), but only for
first 20 modes, $\tau=2958$.}\label{ccortime}
\end{figure}

Thus the coherent structures extracted by SVD have a long lifetime
and produce a slow decay of the temporal autocorrelation function,
defined by
\[
C(\tau) = \langle B(x,y,t+\tau) B(x,y,t) \rangle _{x,y,t},
\]
as shown in Fig. \ref{ccortime}. While small-scale structures
rapidly appear and disappear, the large scale ones evolve slowly.
In that sense, they are truely coherent structures.

\subsection{Magnetic field entropy}

\begin{figure}
%\resizebox{\hsize}{!}{\includegraphics{geisel1.eps}}
\centering
\includegraphics[width=8cm]{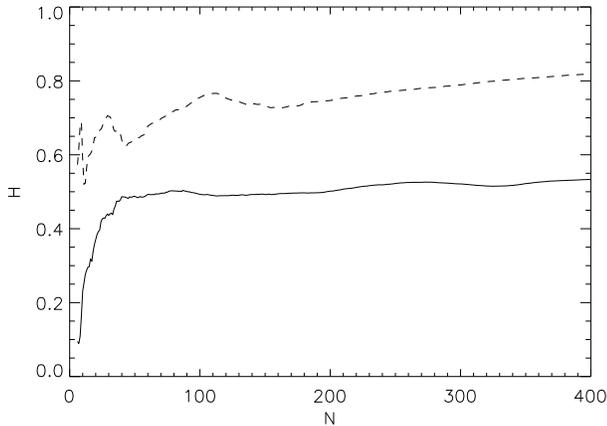}
\caption{Partial entropies $H_N$ as a function of the grid size
$N$. The continuous line is for the subdiffusive source, and the
dashed one for the random source. This entropy is normalized so
that $H=1$ corresponds to maximum disorder.}\label{HN}
\end{figure}

Quantitatively, the coherence degree of the magnetic field can be
measured by the spatial entropy defined by Eq. \ref{eq:H} from the
singular values. This definition involves a limit
$N\rightarrow\infty$, but in practice, for large enough grid
sizes, it can be checked that the quantity computed for a $N\times
N$ subset of $B$
\[
H_N = -\frac{1}{\log N} \sum_{k=1}^N E_k \log E_k
\]
converges toward a well-defined limit as $N$ increases. Computing
the entropy $H_N$ for increasing $N$, we have obtained the curves
displayed on Fig. \ref{HN} which show that the entropy already
converges for matrix sizes about $100\times 100$, although it
seems that the convergence is faster for the subdiffusive source
than for the random source. Thus we may conclude that, provided
large enough grids are considered, the entropy $H$ is fairly
independent of the grid size.

The entropy has a monotonous decay in time and converges toward a
finite value in the steady state (see Tab. \ref{tab:Ht}),
indicating the simultaneous decrease of spatial complexity and the
formation of slowly evolving large scale magnetic field
structures.

\begin{table}[t]
\center
\begin{tabular}{r | r r r r}
\hline \hline
 t  & $100$ & $500$ & $20000$ & $30000$ \\
 H & 0.73 & 0.69 & 0.51 & 0.527 \\
 \hline \hline
\end{tabular}
\caption{Variation of the entropy in time, for the subdiffusive
source and reconnection (see also Fig. \ref{geisel}).}
\label{tab:Ht}
\end{table}

The major result here is that the value toward which the entropy
converges in time exhibit significant differences between the
different sources are used, as summarized in Table \ref{tab:H}.

\begin{table}[t]
\center
\begin{tabular}{c c}
\hline \hline
 source type & H \\
random & 0.8 \\ Ulam & 0.78  \\ Geisel & 0.53 \\
 \hline \hline
\end{tabular}
\caption{Entropy in the steady state for various source types, and dissipation
by reconnection.} \label{tab:H}
\end{table}

%%%%%%%%%%%%%%%%%%%%%%%%%%%%%%%%%%%%%%%%%%%%%%%%%%%%%%%%%%%%%%%%%%
\section{Conclusion and discussion}

To study coronal heating due to dissipation of small-scale current
layers, we have performed a statistical analysis of a simple
model. The model was introduced in Krasnoselskikh et al.
\cite{KPL}, and its principal difference with previous ones is
that the system is driven by small scale homogeneously distributed
sources acting on the entire grid for each time step. The idea to
consider small scale sources is motivated by observations by Benz
\& Krucker (\cite{Benz98,Benz99}) that the heating occurs on the
level of the chromosphere, thus the magnetic field structures,
dissipation of which supplies the energy for the heating, are also
of a small scale.

The question addressed in this paper was the following: if the
actual measurements cannot resolve the characteristic scale of the
heating, in what sense are the "macroscopic" observable properties
influenced by the properties of the smaller scale sources ?

To this purpose we have carried out the comparative analysis of
statistical estimations the large scale spatial characteristics of
the magnetic field such as the correlation length, entropy and
most energetic eigenmodes for the different source types that were
used in the model (random, chaotic and intermittent with anomalous
temporal diffusion).

The "noisy" small scales were filtered out in order to study large
scale of the magnetic field. For this purpose we have
reconstructed the magnetic field from eigenmodes given by SVD that
corresponds to most energetic coherent structures. The less
energetic modes that corresponds to noise level were truncated.

The results can be summarized as follows:

The large scale spatial characteristics of the magnetic field such
as the correlation length, entropy and most energetic eigenmodes
depend significantly on both the statistical properties of small
scale magnetic field sources and the dissipation mechanisms.
\begin{itemize}
\item
It was found that the temporal average of the correlation function
is exponential, i.e the correlation length is finite and not
infinite as supposed to be in SOC systems. This length is a little
bit larger for the reconnection dissipation and it depends on the
dissipation threshold also.
\item
With the subdiffusive (Geisel) source and reconnection dissipation
the correlation significantly departs from the exponential.
\item
The Singular Value Decomposition (SVD) allows to extract the most
energetic magnetic field structures, which are essentially larger
than the source size and persist for long times, supporting the
idea that the plasma can organize on large scales while being
driven by small scale sources.
\item
Moreover, the entropy computed from the singular values of the
magnetic field generated by intermittent sources was found to be
much smaller (about 20-30\%) for the subdiffusive source than for
other sources. The most intensive in space and long life
structures are essentially larger in this case also. That indicate
a higher level of organization in the system than in the random
source case.
\end{itemize}

The clear difference of the characteristics of spatial complexity
in the case of Geisel map sources can be explained in the
following way. This deterministic map produces in each cell a
random-like diffusion slower than usual (subdiffusion) of magnetic
field intensity. On the other hand, the dissipation produces a
normal diffusion of the field, i.e. faster magnetic field
relaxation along the spatial grid (on average), and relates the
temporal properties of the source to spatial properties. This
explains why sources with slower diffusion (Geisel) tend  to form
larger scale and longer lived structures than sources with normal
diffusion (random, Ulam).

Thus we have shown in the framework of our model that the large
scale spatial structure of the magnetic field in the solar
atmosphere also contains important statistical information about
the mechanisms of the coronal heating. Such an information can be
extracted by SVD-based techniques, which are readily applicable to
experimental data and can be used in complement to the usual
analysis of radiated energy.

\begin{acknowledgements} O.~Podladchikova is grateful to French
Embassy in Ukraine for the financial support.
\end{acknowledgements}

\end{document}